\documentstyle[12pt] {article}
\newcommand{\beq}{\begin{equation}}
\newcommand{\eeq}{\end{equation}}
\newcommand{\beqa}{\begin{eqnarray}}
\newcommand{\eeqa}{\end{eqnarray}}
\newcommand{\ba}{\begin{array}}
\newcommand{\ea}{\end{array}}

\begin{document}

\begin{center}
{\large \bf Quantum Signature of the Chaos--Order Transition \\
in a Homogeneous SU(2) Yang--Mills--Higgs System}
\footnote{Presented to the VIII International Conference 
on {\it Symmetry Methods in Physics}, 27 July -- 2 August 1997, 
Joint Institute for Nuclear Physics, Dubna (Russia).}
\end{center}

\vskip 1. truecm

\begin{center}
{\bf Luca Salasnich}\footnote{salasnich@math.unipd.it}
\vskip 0.5 truecm
Dipartimento di Matematica Pura ed Applicata, Universit\`a di Padova, \\
Via Belzoni 7, I 35131 Padova, Italy\\
Istituto Nazionale di Fisica Nucleare, Sezione di Padova, \\
Via Marzolo 8, I 35131 Padova, Italy \\
Istututo Nazionale di Fisica della Materia, Unit\'a di Milano, \\
Via Celoria 16, 20133 Milano, Italy \\
\end{center}

\vskip 1. truecm

{\bf Abstract.} We analyze a spatially homogeneous SU(2) 
Yang--Mills--Higgs system both in classical and quantum mechanics. By using 
the Toda criterion of the Gaussian curvature we find 
a classical chaos--order transition as a function of the Higgs vacuum, 
the Yang--Mills coupling constant and the energy of the system. 
Then, we study the nearest--neighbour spacing 
distribution of the energy levels, which shows 
a Wigner--Poisson transition by increasing the value of the Higgs 
field in the vacuum. This transition is a clear quantum signature 
of the classical chaos--order transition of the system. 
 
\vskip 1. truecm

\section{Introduction}
\par
It is well known that the spatially uniform limit 
of scalar electrodynamics and Yang--Mills theory exhibits classical 
chaotic motion [1--8]. Usually the order--chaos transition 
in these systems has been studied 
numerically with Lyapunov exponents and sections of Poincar\`e. 
Less attention has been paid to analytical criteria. 
\par
In this work we study analytically the suppression of classical chaos 
in the spatially homogenous SU(2) Yang--Mills--Higgs (YMH) system 
induced by the Higgs field. Then we analyze the energy fluctuation properties 
of the system, which give a clear quantum signature 
of the classical chaos--order transition of the system [9--13]. 
\par 
The SU(2) YMH system describes the interaction between a scalar Higgs 
field $\phi$ and three non--Abelian Yang--Mills fields $A_{\mu}^a$, 
$a=1,2,3$. The Lagrangian density of the YMH system [14] is given by 
\beq
L={1\over 2}(D_{\mu}\phi )^+(D^{\mu}\phi ) -V(\phi ) 
-{1\over 4}F_{\mu \nu}^{a}F^{\mu \nu a} \; ,
\eeq
where
\beq
(D_{\mu}\phi )=\partial_{\mu}\phi - i g A_{\mu}^b T^b\phi 
\; ,
\eeq
\beq
F_{\mu \nu}^{a}=\partial_{\mu}A_{\nu}^{a}-\partial_{\nu}A_{\mu}^{a}+
g\epsilon^{abc}A_{\mu}^{b}A_{\nu}^{c} \; ,
\eeq
with $T^b=\sigma^b/2$, $b=1,2,3$, generators of the SU(2) algebra, 
and where the potential of the scalar field (the Higgs field) is
\beq
V(\phi )=\mu^2 |\phi|^2 + \lambda |\phi|^4 \; .
\eeq
We work in the (2+1)--dimensional Minkowski space ($\mu =0,1,2$) and 
choose spatially homogeneous Yang--Mills and the Higgs fields
\beq
\partial_i A^a_{\mu} = \partial_i \phi = 0 \; , \;\;\;\; i=1,2
\eeq
i.e. we consider the system in the region in which space fluctuations of 
fields are negligible compared to their time fluctuations. 
\par
In the gauge $A^a_0=0$ and using the real triplet representation for the 
Higgs field we obtain
$$
L={\dot{\vec \phi}}^2 +
{1\over 2}({\dot {\vec A}}_1^2+{\dot {\vec A}}_2^2) 
-g^2 [{1\over 2}{\vec A}_1^2 {\vec A}_2^2 
-{1\over 2} ({\vec A}_1 \cdot {\vec A}_2)^2+
$$
\beq
+({\vec A}_1^2+{\vec A}_2^2){\vec \phi}^2 
-({\vec A}_1\cdot {\vec \phi})^2 -({\vec A}_2 \cdot {\vec \phi})^2 ]
-V( {\vec \phi} ) \; ,
\eeq
where ${\vec \phi}=(\phi^1,\phi^2,\phi^3)$, 
${\vec A}_1=(A_1^1,A_1^2,A_1^3)$ and ${\vec A}_2=(A_2^1,A_2^2,A_2^3)$. 
\par
When $\mu^2 >0$ the potential $V$ has a minimum at $|{\vec \phi}|=0$, 
but for $\mu^2 <0$ the minimum is at 
$$
|{\vec \phi}_0|=\sqrt{-\mu^2\over 4\lambda }=v \; ,
$$
which is the non zero Higgs vacuum. This vacuum is degenerate 
and after spontaneous symmetry breaking the physical vacuum can be 
chosen ${\vec \phi}_0 =(0,0,v)$. If $A_1^1=q_1$, $A_2^2=q_2$ 
and the other components of the Yang--Mills fields are zero, 
in the Higgs vacuum the Hamiltonian of the system reads 
\beq
H={1\over 2}(p_1^2+p_2^2)
+g^2v^2(q_1^2+q_2^2)+{1\over 2}g^2 q_1^2 q_2^2 \; ,
\eeq
where $p_1={\dot q_1}$ and $p_2={\dot q_2}$. Here $w^2=2 g^2v^2$ is the 
mass term of the Yang--Mills fields. This YMH Hamiltonian is 
a toy model for classical non--linear dynamics, with the attractive feature 
that the model emerges from particle physics. In the next sections we 
analyze first the classical chaos--order transition of the YMH system 
and then its connection to the quantal fluctuations of the energy levels. 

\section{Classical transition from chaos to order}

In this paper we study the chaotic behaviour of this YMH system by using 
the Toda criterion of the Gaussian curvature of the potential energy [16]. 
The Toda criterion is based on a local estimation of the rate of separation 
of neighboring trajectories in the classical phase space of the 
model. To obtain the time evolution of a dynamical system with the Hamiltonian 
\beq
H={p_1^2\over 2}+{p_2^2\over 2} + V(q_1, q_2 ) \; , 
\eeq
the following Hamilton equations have to be solved: 
\beq
{d{\bf q}\over dt}={\partial H\over \partial {\bf p} } \; 
, \;\;\;\;\;\; 
{d{\bf p}\over dt}= - {\partial H\over \partial {\bf q}} \; ,
\eeq
where ${\bf q}=(q_1,q_2)$ and ${\bf p}= (p_1,p_2)$. 
The linearized equation of motion for the deviations are
\beq
{d \delta {\bf p} \over dt} = I \delta {\bf p} \; , 
\;\;\;\;\;\; 
{d \delta {\bf q}\over dt} = - S(t) \delta {\bf q} \; , 
\eeq
where $I$ is the $2\times 2$ identity matrix, and 
\beq
S_{ij}(t) = {\partial^2 V\over \partial q_i q_j}|_{{\bf q}={\bf q}(t)} 
\; ,
\eeq
with ${\bf q}(t)$ solution of the Hamilton equations. 
The stability of the dynamical system is then determined by the 
eigenvalues of the $4\times 4$ stability matrix
\beq
\Gamma({\bf q}(t))= 
\left( \begin{array}{cc}  
            0     &   I \\ 
         -S(t)    &   0 
\end{array}  \right) \; .
\eeq
If at least one of the eigenvalues of the stability matrix $\Gamma$ 
is real, then the separation of the trajectories grows exponentially 
and the motion is unstable. 
Imaginary eigenvalues correspond to stable motion. 
\par
To diagonalize the matrix $\Gamma$, we must first solve the 
equation of motion. 
The problem can be significantly simplified by assuming that the 
time dependence can be eliminated by replacement of the time--dependent 
point ${\bf q}(t)$ of configuration space by a 
time--independent coordinate ${\bf q}$, 
i.e. $\Gamma ({\bf q}(t))=\Gamma ({\bf q})$. 
The eigenvalues then are
\beq
\lambda = \pm [-B\pm \sqrt{B^2 - 4 C} ]^{1\over 2} \; ,
\eeq
where 
\beq
B = \big[ {\partial^2 V\over \partial q_1^2} + 
 {\partial^2 V\over \partial q_2^2} \big] \; ,
\eeq
\beq
C = \big[ {\partial^2 V\over \partial q_1^2} 
{\partial^2 V\over \partial q_2^2} - 
({\partial^2 V\over \partial q_1 q_2})^2 \big] \; .
\eeq
Now, if $B>0$ then with $C\geq 0$ the eigenvalues are purely imaginary 
and the motion is stable, while with $C<0$ the pair of eigenvalues 
becomes real and this leads to exponential separation of neighboring 
trajectories, i.e. chaotic motion. The parameter $c$ has the same sign 
as the Gaussian curvature $K_G$ of the potential--energy surface: 
\beq
K_G(q_1,q_2) = { {\partial^2 V\over \partial q_1^2} 
{\partial^2 V\over \partial q_2^2} - 
({\partial^2 V\over \partial q_1 q_2})^2 \over 
\big[ 1+ ({\partial^2 V\over \partial q_1^2})^2 + 
({\partial^2 V\over \partial q_2^2})^2 \big]^2 } \; . 
\eeq
For our YMH system the potential energy is given by 
\beq
V(q_1 ,q_2)=g^2v^2(q_1^2+q_2^2)+{1\over 2}g^2 q_1^2 q_2^2 \; ,
\eeq
and the $\Gamma$ matrix reads
\beq
\Gamma (q_1,q_2) =
\left( \begin{array}{cccc}  
         0             &    0                   & 1  & 0  \\ 
         0             &    0                   & 0  & 1  \\ 
- 2g^2v^2 - g^2 q_2^2  & - 2g^2 q_1 q_2         & 0  & 0  \\ 
- 2g^2 q_1 q_2         & - 2g^2v^2 - g^2 q_1^2  & 0  & 0  
\end{array}  \right) 
\eeq
At low energy, the motion near the minimum of the potential, 
where the Gaussian curvature is positive, is periodic or quasi--periodic and is 
separated from the instability region by a line of zero curvature; 
if the energy is increased, the system will be, for some initial conditions, 
in a region of negative curvature, where the motion is chaotic. 
According to this scenario, the energy $E_c$ of chaos--order transition 
is equal to the minimum value of the line of zero gaussian 
curvature $K(q_1 ,q_2 )$ on the potential--energy surface. 
For our potential the gaussian curvature vanishes at the points 
that satisfy the equation
\beq
(2g^2v^2 +g^2 q_2^2)(2g^2v^2+g^2q_1^2)-4g^4 q_1^2q_2^2=0 \; .
\eeq
It is easy to show that the minimal energy on the 
zero--curvature line is given by:
\beq
E_c=V_{min}(K_G=0,\bar{q_1})=6 g^2 v^4 \; , 
\eeq
and by inverting this equation 
we obtain $v_c=(E /6g^2)^{1/4}$. We conclude that 
there is a order--chaos transition by increasing the energy $E$ 
of the system and a chaos--order transition by increasing 
the value $v$ of the Higgs field in the vacuum (see also [2]). Thus, 
there is only one transition regulated by the unique parameter $E/(g^2v^4)$. 
\par
It is important to point out that 
{\it in general} the curvature 
criterion guarantees only a {\it local instability} [16] 
and should therefore 
be combined with the Poincar\`e sections [17--18]. 
In the paper [19] we used a fourth--order Runge--Kutta 
method to compute numerically the classical trajectories 
and the Poincar\`e sections. The results 
confirm the analytical predictions of the curvature criterion: 
with $E =10$ and $g=1$ we get the critical value of the onset of chaos 
$v_c=(E /6g^2)^{1/4}\simeq 1.14$. 

\section{Quantum chaos and the Wigner distribution}
\par
The energy fluctuation properties of systems with underlying 
classical chaotic behaviour and time--reversal symmetry agree with 
the predictions of the Gaussian Orthogonal Ensemble (GOE) of 
random matrix theory, whereas quantum analogs of classically 
integrable systems display the characteristics 
of the Poisson statistics [9--12]. Some results in this direction 
for field theories have been obtained by 
Halasz and Verbaarschot: they studied the QCD lattice spectra 
for staggered fermions and its connection to random matrix theory [13]. 
\par
In quantum mechanics the generalized coordinates of the YMH system 
satisfy the usual commutation rules $[{\hat q}_k,{\hat p}_l]=i\delta_{kl}$, 
with $k,l=1,2$. Introducing the creation and destruction operators
\beq
{\hat a}_k=\sqrt{\omega \over 2}{\hat q}_k + 
i \sqrt{1\over 2\omega}{\hat p}_k \; ,
\;\;\;\;
{\hat a}_k^+ = \sqrt{\omega \over 2}{\hat q}_k - 
i \sqrt{1\over 2\omega}{\hat p}_k \; ,
\eeq
the quantum YMH Hamiltonian can be written [15] 
\beq
{\hat H}={\hat H}_0 + {1\over 2} g^2 {\hat V} \; ,
\eeq
where
\beq
{\hat H}_0= \omega ({\hat a}_1^+ {\hat a}_1 + {\hat a}_2^+ {\hat a}_2 + 1) \; ,
\eeq
\beq
{\hat V}= {1 \over 4 \omega^2} ({\hat a}_1 +{\hat a}_1^+)^2 
({\hat a}_2 +{\hat a}_2^+)^2 \; ,
\eeq
with $\omega^2 = 2 g^2 v^2$ and $[{\hat a}_k,{\hat a}_l^+] = \delta_{kl}$, 
$k,l=1,2$. 
\par
The most used quantity to study the local fluctuations of the energy levels 
is the spectral statistics $P(s)$. $P(s)$ is 
the distribution of nearest--neighbour spacings 
$s_i=({\tilde E}_{i+1}-{\tilde E}_i)$ 
of the unfolded levels ${\tilde E}_i$. 
It is obtained by accumulating the number of spacings that lie within 
the bin $(s,s+\Delta s)$ and then normalizing $P(s)$ to unit [9--12]. 
\par
For quantum systems whose classical analogs are integrable, 
$P(s)$ is expected to follow the Poisson limit, i.e. 
$P(s)=\exp{(-s)}$. On the other hand, 
quantal analogs of chaotic systems exhibit the spectral properties of 
GOE with $P(s)= (\pi / 2) s \exp{(-{\pi \over 4}s^2)}$, which is the 
so--called Wigner distribution [9--12]. 
The distribution $P(s)$ is the best spectral statistics to analyze 
shorter series of energy levels and 
the intermediate regions between order and chaos. 
\par
Seligman, Verbaarschot and Zirnbauer [20] analyzed 
a class of two--dimensional anharmonic oscillators with 
polynomial perturbation by using the Brody distribution [21] 
\beq
P(s,\omega)=\alpha (\omega +1) s^{\omega} \exp{(-\alpha s^{\omega+1})} \; ,
\eeq
with 
\beq
\alpha = \big( \Gamma [{\omega +2\over \omega+1}] \big)^{\omega +1} \; .
\eeq
This distribution interpolates between the Poisson distribution ($\omega =0$) 
of integrable systems and the Wigner distribution ($\omega =1$) of 
chaotic ones, and thus the parameter $\omega$ can be used as a simple 
quantitative measure of the degree of chaoticity. 

\begin{center}
\begin{tabular}{|cc|} \hline\hline 
Higgs vacuum $v$ & Brody parameter $\omega$ \\ 
\hline
$1.00$ & $0.99$\\ 
$1.05$ & $0.47$\\
$1.10$ & $0.34$\\ 
$1.15$ & $0.12$\\
$1.20$ & $0.01$\\ 
\hline\hline
\end{tabular}
\end{center}

The Table shows the calculated Brody parameter $\omega$ 
of the $P(s)$ distribution for different values of the Higgs vacuum $v$. 
There is Wigner--Poisson transition by increasing the value $v$ 
of the Higgs field in the vacuum. Thus, by using the P(s) distribution 
and the Brody function, it is possible to give a quantitative measure 
of the degree of quantal chaoticity of the system. 
Our numerical calculations show clearly the quantum 
chaos--order transition and its connection to the classical one.  
\par
We compute the energy levels with 
a numerical diagonalization of the truncated matrix of the quantum 
YMH Hamiltonian  in the basis of the harmonic oscillators [22]. 
If $|n_1 n_2>$ is the basis of the occupation numbers of the two 
harmonic oscillators, the matrix elements are
\beq
<n_{1}^{'}n_{2}^{'}|{\hat H}_0|n_{1}n_{2}>= \omega (n_1+n_2+1) 
\delta_{n_{1}^{'}n_{1}} \delta_{n_{2}^{'}n_{2}} \; ,
\eeq
and
$$
<n_{1}^{'}n_{2}^{'}|{\hat V}|n_{1}n_{2}>=
{1 \over 4 \omega^2}
[\sqrt{n_{1}(n_{1}-1)} \delta_{n^{'}_{1}n_{1}-2}
+\sqrt{(n_{1}+1)(n_{1}+2)}\delta_{n^{'}_{1}n_{1}+2}+
$$
\beq
+(2n_{1}+1)\delta_{n^{'}_{1}n_{1}}]
\times[\sqrt{n_2 (n_2-1)}\delta_{n^{'}_2 n_2-2}+ \sqrt{(n_2+1)(n_2+2)}
\delta_{n^{'}_2 n_2+2}+ (2n_2+1)\delta_{n^{'}_2 n_2}] \; .
\eeq
The symmetry of the potential enables us to split 
the Hamiltonian matrix into 4 sub--matrices 
reducing the computer storage required. These sub--matrices are related 
to the parity of the two occupation numbers $n_1$ and $n_2$: 
even--even, odd--odd, even--odd, odd--even. 
The numerical energy levels depend on the dimension of the truncated matrix: 
we compute the numerical levels in double precision 
increasing the matrix dimension until the first 100 levels converge 
within $8$ digits (matrix dimension $1156\times 1156$) [22--23]. 
\par
We use the first $100$ energy levels of the 4 sub--matrices 
to calculate the $P(s)$ distribution. 
In order to remove the secular variation of the level density as a function 
of the energy $E$, for each value of the coupling constant the 
corresponding spectrum is mapped, by a numerical procedure described in [24], 
into one which has a constant level density. 

\section{Conclusions}
\par
The chaotic behaviour of an 
homogenous YMH system has been studied both in classical and quantum 
mechanics. The Gaussian curvature criterion shows that the chaotic behaviour 
is regulated by the unique parameter $E/(g^2v^4)$. 
The YMH system has a order--chaos transition by 
increasing the energy $E$ and a chaos--order 
transition by increasing the value $v$ of the Higgs field in the vacuum. 
\par
The $P(s)$ distribution of the energy 
levels confirms with great accuracy 
the classical chaos--order transition of the YMH system. 
In particular, the Brody parameter $\omega$ shows a Wigner--Poisson transition 
for the $P(s)$ distribution in correspondence to the classical 
chaos--order transition. 

\section*{Acknowledgments}

The author is grateful to Dr. G.S Pogosyan and the organizing 
committee for their kind invitation to the Conference. 

\section*{References}

\begin{description}

\item{\ [1]} G.K. Savvidy, Nucl. Phys. {\bf B 246}, 302 (1984).

\item{\ [2]} A. Gorski, Acta Phys. Pol. {\bf B 15}, 465 (1984).

\item{\ [3]} T. Kawabe and S. Ohta, Phys. Rev. {\bf D 44}, 1274 (1991).

\item{\ [4]} L. Salasnich, Phys. Rev. {\bf D 52}, 6189 (1995). 

\item{\ [5]} T. Kawabe, Phys. Lett. {\bf B 343}, 254 (1995).

\item{\ [6]} L. Salasnich, Mod. Phys. Lett. {\bf A 10}, 3119 (1995).

\item{\ [7]} J. Segar and M.S. Sriram, Phys. Rev. {\bf D 53}, 3976 (1996).

\item{\ [8]} S.G. Matinyan and B. Muller, 
Phys. Rev. Lett. {\bf 78}, 2515 (1997). 

\item{\ [9]} M.C. Gutzwiller, {\it Chaos in Classical and Quantum Mechanics} 
(Springer, Berlin, 1990).

\item{\ [10]} A.M. Ozorio de Almeida, {\it Hamiltonian Systems: Chaos and 
Quantization} (Cambridge University Press, Cambridge, 1990).

\item{\ [11]} K. Nakamura, {\it Quantum Chaos} 
(Cambridge Nonlinear Science Series, Cambridge, 1993).

\item{\ [12]} G. Casati and B.V. Chirikov, {\it Quantum Chaos} 
(Cambridge University Press, Cambridge, 1995).

\item{\ [13]} M.A. Halasz and 
J.J.M. Verbaarschot, Phys. Rev. Lett. {\bf 74}, 3920 (1995).

\item{\ [14]} C. Itzykson and J.B. Zuber, 
{\it Quantum Field Theory} (McGraw--Hill, New York, 1985).

\item{\ [15]} G.K. Savvidy, Phys. Lett. {\bf B 159}, 325 (1985).

\item{\ [16]} M. Toda, Phys. Lett. {\bf A 48}, 335 (1974).

\item{\ [17]} M. Henon, Physica {\bf D 5}, 412 (1982).

\item{\ [18]} G. Benettin, R. Brambilla and 
L. Galgani, Physica {\bf A 87}, 381 (1977).

\item{\ [19]} L. Salasnich, "Quantum Chaos in a Yang--Mills--Higgs System", 
Preprint DMPA/97/11, Univ. of Padova, to be published in 
Modern Physics Letters A. 

\item{\ [20]} T.H. Seligman, J.J.M. Verbaarschot and M.R. Zirnbauer, 
Phys. Rev. lett. {\bf 53}, 215 (1984).

\item{\ [21]} T.A. Brody, Lett. Nuovo Cimento {\bf 7}, 482 (1973).

\item{\ [22]} S. Graffi, V.R. Manfredi and L. Salasnich, 
Mod. Phys. Lett. {\bf B 9}, 747 (1995).

\item{\ [23]} Subroutine F02AAF, The NAG Fortran Library, Mark 14 
(NAG Ltd, Oxford, 1990).

\item{\ [24]} V.R. Manfredi, Lett. Nuovo Cimento {\bf 40}, 135 (1984). 

\end{description}

\end{document}